\documentclass
[superscriptaddress,secnumarabic,amssymb,amsmath,nobibnotes,aps,prd,showkeys,showpacs,nofootinbib,12pt]{revtex4}%
\usepackage{graphicx}
\usepackage{epsf}
\usepackage{bm}
\usepackage{amsmath}
\usepackage{amsfonts}
\usepackage{amssymb}%
\providecommand{\U}[1]{\protect\rule{.1in}{.1in}}
\newcommand{\be}{\begin{equation}}
\newcommand{\ee}{\end{equation}}

\newcommand{\mincir}{\raise
-3.truept\hbox{\rlap{\hbox{$\sim$}}\raise4.truept\hbox{$<$}\ }}
\newcommand{\magcir}{\raise
-3.truept\hbox{\rlap{\hbox{$\sim$}}\raise4.truept\hbox{$>$}\ }}

\begin{document}
\title{Analytic solution and Noether symmetries for the hyperbolic inflationary model
in the Jordan frame}
\author{Andronikos Paliathanasis}
\email{anpaliat@phys.uoa.gr}
\affiliation{Institute of Systems Science, Durban University of Technology, Durban 4000,
South Africa }
\affiliation{Instituto de Ciencias F\'{\i}sicas y Matem\'{a}ticas, Universidad Austral de
Chile, Valdivia 5090000, Chile}

\begin{abstract}
The Noether symmetry analysis is applied for the study of a multifield
cosmological model in a spatially flat FLRW background geometry. The
gravitational Action Integral consists of two scalar fields, the Brans-Dicke
field and a second scalar field minimally coupled to gravity. However, the two
scalar fields interact in the kinetic terms. This multifield has been found to
describe the equivalent of hyperbolic inflation in the Jordan frame. The
application of Noether's theorems constrain the free parameters of the model
that conservation laws exist. We find that the field equations form an
integrable dynamical system and the analytic solution is derived.

\end{abstract}
\maketitle

\section{Introduction}

\label{sec1}

Scalar fields play an important role in the description of the cosmological
evolution \cite{sot1}. With the introduction of the scalar fields in the
Einstein-Hilbert Action the new degrees of freedom drive the dynamics of the
cosmological parameters such that they explain the cosmological observations
\cite{Teg,Kowal}. The quintessence scalar field model\ is very simple model
which describes the so-called dark energy and is responsible for the late-time
acceleration phase of the universe \cite{quin1,quin2}. On the hand, inflation
\cite{inf1} has been proposed to solve the flatness, the horizon and the
isotropization problems. Inflation describes a very rapid acceleration phase
during the early stage of the universe and it is attributed to the inflaton
field \cite{inf2,inf3,inf4}. There is a plethora of proposed scalar field
models in the literature, see for instance
\cite{sf1,sf2,sf3,sf4,sf5,sf6,sf7,sf8} and references therein. \ 

In the middle of the previous century, Brans and Dicke \cite{Brans} proposed a
gravitational model with a scalar field which satisfies Mach's principle.
Indeed, the existence of the scalar field is essential for the physical space
and the scalar field interacts with the gravity in the Action Integral, that
is, the scalar field is nonminimally coupled to gravity. Generalizations of
the Brans-Dicke model are known as scalar-tensor theories \cite{sf9}. In
\cite{sf10}, Hordenski derived the most general Action Integral for the
scalar-tensor theory. Brans-Dicke theory is defined on the Jordan frame
\cite{Jord}, while the gravitational model depends upon a free parameter known
as the Brans-Dicke parameter. The Brans-Dicke model has been used as a model
for the description of the dark energy \cite{f1} and of the inflationary epoch
\cite{f2,f3,f4}.

A two-scalar field model which has drawn the attention of cosmologists in
recent years is the Chiral model \cite{ch1,ch2,ch3}. In the Chiral model, the
two scalar fields are minimally coupled to gravity, that is, they are defined
in the Jordan frame. However, the two scalar field interact in the kinetic
term. Specifically, from the kinetic components of the scalar fields we can
define a second-rank tensor, which for the Chiral model is a two-dimensional
hyperbolic sphere. Thus, there is no \textquotedblleft
coordinate\textquotedblright\ system where there is no interaction for the two
fields. This is in contrast to the quintom model in which the kinetic
components define the two-dimensional flat space \cite{quintom1}. This
specific two-scalar field model has been widely studied in the literature
\cite{ch01,ch02,ch03} and various extensions for which the scalar fields may
have negative energy density have been proposed before \cite{ch04,ch05}.

For a specific potential function, the Chiral model provides a very
interesting scaling solution in which the two scalar fields contribute to the
cosmological fluid \cite{vr91,vr92}. The scaling solution describes
acceleration and the solution is described as hyperbolic inflation or
hyperinflation. Because of the existence of the second scalar field in the
hyperinflation, the curvature perturbations depend upon the number of e-fold
\cite{in1}, while the initial conditions at the start and in the end of the
inflation can be different \cite{in1}, while the non-Gaussianities in the
power spectrum are supported by this model \cite{in2}. Recently, in
\cite{pal1} a multi-scalar field model has been proposed consisting of
two-scalar fields, the Brans-Dicke field and a second field which is coupled
to the Brans-Dicke field in the kinetic term, but it is minimally coupled to
gravity. The latter model is defined in the Jordan frame. However, under a
conformal map the equivalent Action Integral in the Einstein frame is that of
the Chiral theory. The extended Brans-Dicke theory admits a scaling asymptotic
solution which has similar dynamical properties with the hyperbolic
inflationary solution for the Chiral model. Indeed, the asymptotic solution
describes inflation, in which the two scalar fields contribute to the
cosmological solution, while this specific solution corresponds to a spiral
attractor. This dynamical property remains invariant for the two models under
the conformal transformation which relates the two theories.

In this piece of study, we investigate the conservation laws and the
integrability properties for the hyperbolic inflationary model in the Jordan
frame. For the purposes of this study we make use of the property that the
gravitational field equations admit a minisuperspace description so that the
Noether symmetry analysis \cite{ns1} can be applied. Noether's theorems
provide a systematic approach for the determination of infinitesimal
transformations which leave invariant the variational principle. Moreover, the
generators of the invariant infinitesimal transformations can be used in a
simple way to construct conservation laws. Because of the simplicity of the
applications of Noether's theorems and of the importance of the given results,
Noether symmetries have been the subject of study in various gravitational
systems \cite{ns2,ns3,ns4,ns5,ns6}. The plan of the paper is as follows.

In Section \ref{sec2} we present the cosmological model of our consideration
and we derive the minisuperspace and the point-like Lagrangian which describes
the field equations. The basic properties and definitions for the theories of
point transformations are given in Section \ref{sec3}. Moreover, we find the
Noether symmetries and we construct the corresponding conservation laws for
the field equations. In Section \ref{sec4} we determine the analytic solution
for the cosmological model of our analysis.\ We define canonical variables and
we derive the analytic solution. Finally, in \ref{conc} we summarize our results.

\section{Field equations}

\label{sec2}

The cosmological model of our consideration is that of a spatially flat
Friedmann--Lema\^{\i}tre--Robertson--Walker (FLRW) geometry described by the
line element
\begin{equation}
ds^{2}=dt^{2}-a^{2}\left(  t\right)  \left(  dx^{2}+dy^{2}+dz^{2}\right)  ,
\label{bd.05}%
\end{equation}
where $a\left(  t\right)  $ is the scale factor.

The field equations follow from the variation of the Action Integral%
\begin{equation}
S_{A}=\int dx^{4}\sqrt{-g}\left[  \frac{1}{2}\phi R+\frac{1}{2}\frac
{\omega_{BD}}{\phi}g^{\mu\nu}\phi_{;\mu}\phi_{;\nu}+\frac{1}{2}F^{2}\left(
\phi\right)  g^{\mu\nu}\psi_{;\mu}\psi_{;\nu}+V\left(  \phi\right)  \right]  ,
\label{bb.01}%
\end{equation}
where $\phi\left(  x^{\kappa}\right)  $ is the Brans-Dicke field, $\omega
_{BD}$ is the Brans-Dicke parameter, $V\left(  \phi\left(  x^{\kappa}\right)
\right)  $ is the potential function and $\psi\left(  x^{\kappa}\right)  $ is
the second scalar field minimally coupled to gravity.

For the line element (\ref{bd.05}) and the Action Integral (\ref{bb.01}) we
derive the field equations
\begin{equation}
-3H^{2}=3H\frac{\dot{\phi}}{\phi}-\frac{\omega_{BD}}{2}\left(  \frac{\dot
{\phi}}{\phi}\right)  ^{2}-\frac{1}{2}\frac{F^{2}\left(  \phi\right)  }{\phi
}\dot{\psi}^{2}-\frac{1}{\phi}V\left(  \phi\right)  ~, \label{con1}%
\end{equation}%
\begin{equation}
-\left(  3\phi H^{2}+2\phi\dot{H}\right)  =2H\dot{\phi}+\frac{\omega_{BD}%
}{2\phi}\dot{\phi}^{2}+\frac{1}{2}F^{2}\left(  \phi\right)  \dot{\psi}%
^{2}+\ddot{\phi}-V\left(  \phi\right)
\end{equation}%
\begin{equation}
\omega_{BD}\left(  \ddot{\phi}-\frac{1}{2}\left(  \frac{\dot{\phi}}{\phi
}\right)  ^{2}+3H\dot{\phi}\right)  +6H^{2}\phi+\phi\left(  3\dot{H}+V_{,\phi
}-\frac{1}{2}\left(  F^{2}\right)  _{,\phi}\dot{\psi}^{2}\right)  =0~,
\end{equation}%
\begin{equation}
\ddot{\psi}+3H\dot{\psi}+\left(  \ln\left(  F^{2}\right)  \right)  _{,\phi
}\dot{\phi}\dot{\psi}=0,
\end{equation}
in which we have assumed that the scalar fields inherit the symmetries of the
background space, that is $\phi\left(  x^{\kappa}\right)  =\phi\left(
t\right)  $, $\psi\left(  x^{\kappa}\right)  =\psi\left(  t\right)  $,
$H=\frac{\dot{a}}{a}$ is the Hubble function and $\dot{a}=\frac{da}{dt}$.

It is easy to see that the cosmological field equations follow from the
variation of the point-like Lagrangian function%
\begin{equation}
\mathcal{L}\left(  a,\dot{a},\phi,\dot{\phi},\psi,\dot{\psi}\right)
=3a\phi\dot{a}^{2}+3a^{2}\dot{a}\dot{\phi}-\frac{\omega_{BD}}{2\phi}a^{3}%
\dot{\phi}^{2}-\frac{1}{2}a^{3}F^{2}\left(  \phi\right)  \dot{\psi}^{2}%
+a^{3}V\left(  \phi\right)  ~. \label{ln.01}%
\end{equation}
The constraint equation (\ref{con1}) can be seen as Hamiltonian constraint for
the autonomous dynamical system.

From the point-like Lagrangian we define the minisuperspace which has
dimension three and line element%
\begin{equation}
ds_{\gamma}^{2}=6a\phi da^{2}+6a^{2}dad\phi-\frac{\omega_{BD}}{\phi}a^{3}%
d\phi^{2}-a^{3}F^{2}\left(  \phi\right)  d\psi^{2}.
\end{equation}

Hyperbolic inflation in the Jordan frame is recovered when $F\left(
\phi\right)  =F_{0}\phi^{\kappa}$ and $V\left(  \phi\right)  =V_{0}%
\phi^{\lambda}$. Hence, these two functions are considered in the following Section.

\section{Noether symmetries and conservation laws}

\label{sec3}

We review the basic definitions concerning invariant point transformations and
Noether symmetries of systems of second-ordinary differential equations.

Assume the dynamical system%
\begin{equation}
\mathbf{\ddot{y}}=\omega\left(  t,\mathbf{y,\dot{y}}\right)  . \label{Lie.0}%
\end{equation}
Then a vector field
\begin{equation}
X=\xi\left(  t,\mathbf{y}\right)  \partial_{t}+\mathbf{\eta}\left(
t,\mathbf{y}\right)  \partial_{\mathbf{y}}%
\end{equation}
\ in the augmented space $\{t,x^{i}\}$ is a point symmetry of the system of
differential equations (\ref{Lie.0})\ if the following condition is satisfied
\cite{ns6}
\begin{equation}
X^{\left[  2\right]  }\left(  \mathbf{\ddot{y}}-\omega\left(  t,\mathbf{y,\dot
{y}}\right)  \right)  =0, \label{Lie.1}%
\end{equation}
where $X^{\left[  2\right]  }$\ is the second prolongation of $X$ defined as
follows%
\begin{equation}
X^{\left[  2\right]  }=\xi\partial_{t}+\mathbf{\eta}\partial_{\mathbf{y}%
}+\left(  \mathbf{\dot{\eta}}-\mathbf{\dot{y}}\dot{\xi}\right)  \partial
_{\mathbf{\dot{y}}}+\left(  \mathbf{\ddot{\eta}}-\mathbf{\dot{y}}\ddot{\xi
}-2\mathbf{\ddot{y}}\dot{\xi}\right)  \partial_{\mathbf{\ddot{y}}}.
\label{Lie.2}%
\end{equation}

Thus, if $X$ is a symmetry vector, then under the infinitesimal transformation%
\begin{equation}
t^{\prime}=t+\varepsilon\xi\left(  t,\mathbf{y}\right)  ~,~y^{\prime
}=y+\varepsilon\mathbf{\eta}\left(  t,\mathbf{y}\right)  ~,
\end{equation}
the dynamical system (\ref{Lie.0}) remains invariant which means that
trajectories of solutions drive along to the vector field $X$.

Condition (\ref{Lie.1}) is equivalent to the relation
\begin{equation}
\left[  X^{\left[  1\right]  },A\right]  =\lambda\left(  x^{a}\right)  A,
\label{Lie.3a}%
\end{equation}
where $X^{\left[  1\right]  }~$is the first prolongation of $X$ and $A$ is the
Hamiltonian vector field%
\begin{equation}
A=\partial_{t}+\mathbf{\dot{y}}\partial_{\mathbf{y}}+\omega\left(
t,\mathbf{y,\dot{y}}\right)  \partial_{\mathbf{y}}. \label{Lie.4}%
\end{equation}

If the system of differential equations results from a first-order Lagrangian
$\mathcal{L}=\mathcal{L}\left(  t,\mathbf{y,\dot{y}}\right)  ,$ then a Lie
symmetry $X$ of the system is a Noether symmetry of the Lagrangian if the
additional condition
\begin{equation}
X^{\left[  1\right]  }\mathcal{L}+\mathcal{L}\frac{d\xi}{dt}=\frac{df}{dt}
\label{Lie.5}%
\end{equation}
is satisfied, where $f=f\left(  t,\mathbf{y}\right)  $\ is a boundary function
and%
\begin{equation}
X^{\left[  1\right]  }=\xi\partial_{t}+\mathbf{\eta}\partial_{\mathbf{y}%
}+\left(  \mathbf{\dot{\eta}}-\mathbf{\dot{y}}\dot{\xi}\right)  \partial
_{\mathbf{\dot{y}}}.
\end{equation}

According to Noether's second theory to every symmetry there corresponds a
first integral (a Noether integral) of the system of equations (\ref{Lie.0})
which is given by the formula:%
\begin{equation}
I\left(  X\right)  =\xi E_{H}-\frac{\partial\mathcal{L}}{\partial
\mathbf{\dot{y}}}\mathbf{\eta}+f, \label{Lie.6}%
\end{equation}
where $E_{H}\left(  t,\mathbf{y,\dot{y}}\right)  $ is the Hamiltonian function
of $\mathcal{L}\left(  t,\mathbf{y,\dot{y}}\right)  $.

Consider now the infinitesimal transformation%
\begin{equation}
t^{\prime}=t+\varepsilon\xi\left(  t,a,\phi,\psi\right)  ~,~a^{\prime
}=a+\varepsilon\eta^{a}\left(  t,a,\phi,\psi\right)  ~,
\end{equation}%
\begin{equation}
\phi^{\prime}=\phi+\varepsilon\eta^{\phi}\left(  t,a,\phi,\psi\right)
~,~\psi^{\prime}=\psi+\varepsilon\eta^{\psi}\left(  t,a,\phi,\psi\right)  ~,
\end{equation}
and generator the vector field
\begin{equation}
X=\xi\partial_{t}+\eta^{a}\partial_{a}+\eta^{\phi}\partial_{\phi}+\eta^{\psi
}\partial_{\psi}.
\end{equation}

Then, the application of Noether's condition (\ref{Lie.5}) for the Lagrangian
function (\ref{ln.01}) with $F\left(  \phi\right)  =F_{0}\phi^{\kappa}$ and
$V\left(  \phi\right)  =V_{0}\phi^{\lambda}$, gives a system of differential
equations which constrain the infinitesimal parameters. The results are
summarized in the following propositions.

\textbf{Proposition 1:}\textit{ The point-like Lagrangian (\ref{ln.01}) with
}$F\left(  \phi\right)  =F_{0}\phi^{\kappa}$\textit{ and }$V\left(
\phi\right)  =V_{0}\phi^{\lambda}$\textit{ for arbitrary values of the
parameters }$\kappa,\lambda$\textit{ admits the Noether symmetries }%
$X_{1}=\partial_{t},~X_{2}=\partial_{\psi}$\textit{. However, when }%
$\kappa=\frac{1}{2}$\textit{, }$\lambda=1$\textit{ there exist the additional
symmetry vectors }$X_{3}=a\partial_{a}-3\phi\partial_{\phi}$\textit{,~}%
$X_{4}=-\frac{a}{3}\psi\partial_{a}+\phi\psi\partial_{\phi}+\frac{2}{F_{0}%
^{2}}\ln\left(  \frac{a}{\phi^{1+\omega_{BD}}}\right)  \partial_{\psi}%
;~$\textit{while for }$\lambda=2\kappa$\textit{, }$\kappa=\frac{\sqrt{3\left(
3+2\omega_{BD}\right)  }}{4}+\frac{3}{4}$\textit{ the field equations admit
the extra Noether symmetries }$\bar{X}_{3}=\phi^{\beta_{1}}a^{\beta_{2}%
}\left(  \partial_{a}-\frac{6}{\sqrt{3\left(  3+2\omega_{BD}\right)  }}%
\frac{\phi}{a}\partial_{\phi}\right)  $\textit{ where }$\beta_{1}=\frac
{\sqrt{3}\left(  \omega_{BD}+1\right)  }{2\sqrt{\left(  3+2\omega_{BD}\right)
}}-\frac{1}{2}$\textit{ }$,~\beta_{2}=-\frac{1}{2}-\frac{3}{2\sqrt{3\left(
3+2\omega_{BD}\right)  }}$ and $\bar{X}_{4}=\phi^{\beta_{1}}a^{\beta_{2}%
}\left(  -\psi\partial_{\alpha}+\frac{\psi\phi}{a}\partial_{\phi}%
+\frac{8\left(  \omega_{BD}+\frac{3}{2}\right)  }{F_{0}^{2}\left(
\sqrt{3\left(  3+2\omega_{BD}\right)  }+1\right)  a}\phi^{-\frac
{1+\sqrt{3\left(  3+2\omega_{BD}\right)  }}{2}}\partial_{\psi}\right)  $.
Moreover, for arbitrary value of $\kappa$ and $\lambda=1$ there exist the
Noether symmetry vector $X_{5}=-\frac{a}{3}\partial_{a}+\phi\partial_{\phi
}-\frac{\sqrt{3\left(  3+2\omega_{BD}\right)  }+1}{4F_{0}^{2}}\psi
\partial_{\psi}$.

\textbf{Proposition 2:}\textit{ According to Noether' second theorem and for
the expression (\ref{Lie.6}) the cosmological model of our consideration
admits the conservation laws~}$I\left(  X_{1}\right)  =E_{H}$\textit{,
}$I\left(  X_{2}\right)  =a^{3}F_{0}^{2}\phi^{2\kappa}\dot{\psi}$\textit{ for
arbitrary values of the free parameters }$\kappa,\lambda$\textit{. For
}$\left(  \kappa,\lambda\right)  =\left(  \frac{1}{2},1\right)  $\textit{ the
additional conservation laws are }%
\begin{equation}
I\left(  X_{3}\right)  =a\left(  6a\phi\dot{a}+3a^{2}\dot{\phi}\right)
-3\phi\left(  3a^{2}\dot{a}-\frac{\omega_{BD}}{\phi}a^{3}\dot{\phi}\right)  ~,
\end{equation}%
\begin{equation}
I\left(  X_{4}\right)  =-\frac{a}{3}\psi\left(  6a\phi\dot{a}+3a^{2}\dot{\phi
}\right)  +\phi\psi\left(  3a^{2}\dot{a}-\frac{\omega_{BD}}{\phi}a^{3}%
\dot{\phi}\right)  ~+2a^{3}\phi^{2\kappa}\dot{\psi}\ln\left(  \frac{a}%
{\phi^{1+\omega_{BD}}}\right)  .
\end{equation}
\textit{For }$\left(  \kappa,\lambda\right)  =\left(  \kappa,2\kappa\right)
$\textit{, }$\kappa=\frac{\sqrt{3\left(  3+2\omega_{BD}\right)  }}{2}+\frac
{3}{4}$\textit{, the corresponding conservation laws are}%
\begin{equation}
I\left(  \bar{X}_{3}\right)  =\phi^{\beta_{1}}a^{\beta_{2}}\left(  \left(
6a\phi\dot{a}+3a^{2}\dot{\phi}\right)  -\frac{6}{\sqrt{3\left(  3+2\omega
_{BD}\right)  }}\phi\left(  3a\dot{a}-\frac{\omega_{BD}}{\phi}a^{2}\dot{\phi
}\right)  \right)  ,
\end{equation}%
\begin{equation}
I\left(  \bar{X}_{4}\right)  =\phi^{\beta_{1}}a^{\beta_{2}}\left(
-\psi\left(  6a\phi\dot{a}+3a^{2}\dot{\phi}\right)  +\psi\phi\left(  3a\dot
{a}-\frac{\omega_{BD}}{\phi}a^{2}\dot{\phi}\right)  +\frac{8\left(
\omega_{BD}+\frac{3}{2}\right)  a^{2}\dot{\psi}}{\left(  \sqrt{3\left(
3+2\omega_{BD}\right)  }+1\right)  }\phi^{2\kappa-\frac{1+\sqrt{3\left(
3+2\omega_{BD}\right)  }}{2}}\right)  .
\end{equation}
\textit{Finally, for arbitrary }$\kappa$\textit{ and }$\lambda=1$\textit{, the
additional conservation law is }%
\begin{equation}
I\left(  X_{5}\right)  =-\frac{a}{3}\left(  6a\phi\dot{a}+3a^{2}\dot{\phi
}\right)  +\phi\left(  3a^{2}\dot{a}-\frac{\omega_{BD}}{\phi}a^{3}\dot{\phi
}\right)  -\frac{\sqrt{3\left(  3+2\omega_{BD}\right)  }+1}{4}\phi^{2\kappa
}\psi\dot{\psi}.
\end{equation}

We observe that the set of the conservation laws $\left(  I\left(
X_{1}\right)  ,\mathit{~}I\left(  X_{2}\right)  ,\mathit{~}I\left(
X_{3}\right)  \right)  $ and $\left(  I\left(  X_{1}\right)  ,\mathit{~}%
I\left(  X_{2}\right)  ,\mathit{~}I\left(  \bar{X}_{3}\right)  \right)  $ are
independent and in involution, that is $\left\{  I\left(  X_{A}\right)
,I\left(  X_{B}\right)  \right\}  =0$, $A,B=1,2,3$ and $\ \left\{  ,\right\}
$ is the Poisson bracket. Consequently, according to Liouville's theorem the
field equations of this two-dimensional system are integrable. Specifically,
because they admit additional conservation laws, they are superintegrable
\cite{arn1}. For these two cases we proceed with the derivation of the
analytic solutions.

\section{Analytic solutions}

\label{sec4}

The procedure that we apply for the derivation of the analytic solutions is
summarized in the following steps. For the vector fields $X_{A}$ we find the
normal variables by solving the system of differential equations
\begin{equation}
X_{A}\left(  F\left(  a,\phi,\psi\right)  \right)  =0.
\end{equation}
We write the field equations in the new coordinates and we solve the resulting system.

\subsection{Model A}

For the first case of our analysis $\left(  \kappa,\lambda\right)  =\left(
\frac{1}{2},1\right)  $, and the symmetry vector $X_{3}$ we determine the
normal coordinates $\left(  a,\Phi,\psi\right)  $ where%
\begin{equation}
\phi=\frac{\Phi}{a^{3}}.
\end{equation}

Thus the point-like Lagrangian (\ref{ln.01}) is%
\begin{equation}
\mathcal{L}\left(  a,\dot{a},\Phi,\dot{\Phi},\psi,\dot{\psi}\right)  =\frac
{3}{2}\left(  4+3\omega_{BD}\right)  \Phi\left(  \frac{\dot{a}}{a}\right)
^{2}-3\left(  1+\omega_{BD}\right)  \left(  \frac{\dot{a}}{a}\right)
\dot{\Phi}+\frac{\omega_{BD}}{2}\frac{\dot{\Phi}^{2}}{\Phi}+\frac{1}{2}%
F_{0}^{2}\Phi\dot{\psi}^{2}-V_{0}\Phi. \label{sc.01}%
\end{equation}

Consequently the field equations are
\begin{equation}
\frac{3}{2}\left(  4+3\omega_{BD}\right)  \Phi\left(  \frac{\dot{a}}%
{a}\right)  ^{2}-3\left(  1+\omega_{BD}\right)  \left(  \frac{\dot{a}}%
{a}\right)  \dot{\Phi}+\frac{\omega_{BD}}{2}\frac{\dot{\Phi}^{2}}{\Phi}%
+\frac{1}{2}F_{0}^{2}\Phi\dot{\psi}^{2}+V_{0}\Phi=0,
\end{equation}%
\begin{equation}
\left(  4+3\omega_{BD}\right)  \left(  \Phi\ddot{a}+\dot{a}\dot{\Phi}%
-\Phi\frac{\dot{a}^{2}}{a}\right)  -\left(  1+\omega_{BD}\right)  a\ddot{\Phi
}=0~,
\end{equation}%
\begin{equation}
3\left(  2+\omega_{BD}\right)  \dot{a}^{2}+6\left(  1+\omega_{BD}a\right)
\ddot{a}+a^{2}\left(  \omega_{BD}\left(  \frac{\dot{\Phi}^{2}}{\Phi^{2}%
}-2\frac{\ddot{\Phi}}{\Phi}\right)  +\left(  F_{0}^{2}\dot{\psi}^{2}%
-2V_{0}\right)  \right)  =0~.
\end{equation}%
\begin{equation}
\ddot{\psi}+\frac{\dot{\Phi}}{\Phi}\dot{\psi}=0.
\end{equation}

Consequently, with the use of the constraint equation we write
\begin{equation}
I\left(  X_{2}\right)  =\Phi\dot{\psi}\,,~
\end{equation}%
\begin{equation}
\ddot{\Phi}=\frac{\left(  1+V_{0}\right)  \left(  4+3\omega_{BD}\right)
}{3+2\omega_{BD}}\Phi~, \label{sc.01aa}%
\end{equation}%
\begin{equation}
\ddot{a}=\left(  1+V_{0}\right)  \left(  \frac{1+\omega_{BD}}{3+2\omega_{BD}%
}\right)  a+\frac{\dot{a}^{2}}{a}-\frac{\dot{\Phi}}{\Phi}\dot{a}~,
\end{equation}
or equivalently%
\begin{equation}
\dot{H}=\left(  1+V_{0}\right)  \left(  \frac{1+\omega_{BD}}{3+2\omega_{BD}%
}\right)  -\frac{\dot{\Phi}}{\Phi}H. \label{sc.01bb}%
\end{equation}

Thus,
\begin{equation}
\dot{H}\Phi+\dot{\Phi}H-\left(  1+V_{0}\right)  \left(  \frac{1+\omega_{BD}%
}{3+2\omega_{BD}}\right)  \Phi=0,
\end{equation}
where by replacing from (\ref{sc.01aa}) we find%
\begin{equation}
\left(  H\Phi\right)  ^{\cdot}-\left(  \frac{1+\omega_{BD}}{4+3\omega_{BD}%
}\right)  \ddot{\Phi}=0,
\end{equation}
that is, the conservation law it follows%
\begin{equation}
\left(  H\Phi\right)  -\left(  \frac{1+\omega_{BD}}{4+3\omega_{BD}}\right)
\dot{\Phi}=\bar{I}_{0}%
\end{equation}
where $\bar{I}_{0}$ is an integration constant. The latter expresion is
analogue to the Noetherian conservation law $I\left(  X_{3}\right)  $.

Hence, for the scalar field the analytic solution follows%
\begin{equation}
\Phi\left(  t\right)  =\Phi_{1}e^{\Omega t}+\Phi_{2}e^{-\Omega t}%
~,~\Omega=\sqrt{\frac{\left(  1+V_{0}\right)  \left(  4+3\omega_{BD}\right)
}{3+2\omega_{BD}}}.\text{ }%
\end{equation}

For initial conditions, for which $\Phi_{1}\Phi_{2}=0$, the analytic solution
for the Hubble function is
\begin{equation}
H\left(  t\right)  =\mp\frac{\sqrt{1+V_{0}}\left(  1+\omega_{BD}\right)
}{\sqrt{\left(  3+2\omega_{BD}\right)  \left(  4+3\omega_{BD}\right)  }}%
+H_{0}e^{\pm\Omega t},
\end{equation}
respectively. Therefore, the scalar factor is derived to be
\begin{equation}
\ln a\left(  t\right)  =\mp\frac{\sqrt{1+V_{0}}\left(  1+\omega_{BD}\right)
}{\sqrt{\left(  3+2\omega_{BD}\right)  \left(  4+3\omega_{BD}\right)  }}%
t\pm\frac{1}{\Omega}e^{\pm\Omega t}.
\end{equation}

The effective equation of state parameter $w_{eff}=-1-\frac{2}{3}\frac{\dot
{H}}{H^{2}}$ is calculated.%
\begin{equation}
w_{eff}\left(  t\right)  =-1\mp\frac{2}{3}\frac{\Omega e^{\pm\Omega t}H_{0}%
}{\left(  \mp\frac{\sqrt{1+V_{0}}\left(  1+\omega_{BD}\right)  }{\sqrt{\left(
3+2\omega_{BD}\right)  \left(  4+3\omega_{BD}\right)  }}+H_{0}e^{\pm\Omega
t}\right)  ^{2}}.
\end{equation}
For $H_{0}=0$, it is easy to observe that the de Sitter Universe,
\begin{equation}
\ln a\left(  t\right)  =\mp\frac{\sqrt{1+V_{0}}\left(  1+\omega_{BD}\right)
}{\sqrt{\left(  3+2\omega_{BD}\right)  \left(  4+3\omega_{BD}\right)  }}t~,
\end{equation}
is recovered. However, for $H_{0}\neq0$ and for large values of $t$, the de
Sitter universe is the asymptotic solution.

In general for $\Phi_{1}\Phi_{2}\neq0$, the Hubble function is derived to be%
\begin{equation}
H\left(  t\right)  =\frac{\sqrt{1+V_{0}}}{\sqrt{\left(  3+2\omega_{BD}\right)
\left(  4+3\omega_{BD}\right)  }}\frac{1}{\left(  \Phi_{1}e^{2\Omega t}%
+\Phi_{2}\right)  }+\frac{H_{0}}{\Phi_{1}e^{2\Omega t}+\Phi_{2}}\exp\left(
\Omega t\right)  .~
\end{equation}

Easily we observe that the $w_{eff}\left(  t\right)  $ for the latter solution
for large values of $t$, reaches asymptotically the value $w_{eff}\left(
t\right)  \rightarrow-1$, thus, the de Sitter Universe is an asymptotic
solution for the dynamical system.

\subsection{Model B}

For the second model of our analysis, that is, for $\left(  \kappa
,\lambda\right)  =\left(  \kappa,2\kappa\right)  $\textit{, }$\kappa
=\frac{\sqrt{3\left(  3+2\omega_{BD}\right)  }}{2}+\frac{3}{4}$, the normal
coordinates are $\left(  a,\Xi,\psi\right)  $, in which
\begin{equation}
\Xi=\phi a^{\frac{6}{\sqrt{3\left(  3+2\omega_{BD}\right)  }+3}}.
\end{equation}
In the new variables the point-like Lagrangian is%
\begin{align}
\mathcal{L}\left(  a,\dot{a},\Xi,\dot{\Xi},\psi,\dot{\psi}\right)   &
=a^{2-\frac{6}{3+\sqrt{3\left(  3+2\omega_{BD}\right)  }}}\left(
\sqrt{3\left(  3+2\omega_{BD}\right)  }\dot{a}\dot{\Xi}-\frac{\omega_{BD}}%
{2}\frac{a}{\Xi}\dot{\Xi}^{2}\right) \nonumber\\
&  -\frac{1}{2}F_{0}^{2}\Xi^{\frac{3+\sqrt{3\left(  3+2\omega_{BD}\right)  }%
}{2}}\dot{\psi}^{2}+V_{0}\Xi^{\frac{3+\sqrt{3\left(  3+2\omega_{BD}\right)  }%
}{2}}.
\end{align}

Hence, the field equations are%
\begin{equation}
a^{2-\frac{6}{3+\sqrt{3\left(  3+2\omega_{BD}\right)  }}}\left(
\sqrt{3\left(  3+2\omega_{BD}\right)  }\dot{a}\dot{\Xi}-\frac{\omega_{BD}}%
{2}\frac{a}{\Xi}\dot{\Xi}^{2}\right)  -\frac{1}{2}F_{0}^{2}\Xi^{\frac
{3+\sqrt{3\left(  3+2\omega_{BD}\right)  }}{2}}\dot{\psi}^{2}+V_{0}\Xi
^{\frac{3+\sqrt{3\left(  3+2\omega_{BD}\right)  }}{2}}=0,
\end{equation}%
\begin{equation}
\ddot{\Xi}=\frac{\sqrt{3\left(  3+2\omega_{BD}\right)  }-3+\omega_{BD}\left(
\sqrt{3\left(  3+2\omega_{BD}\right)  }-2\right)  }{2\left(  3+2\omega
_{BD}\right)  }\frac{\dot{\Xi}^{2}}{\Xi}~, \label{s11}%
\end{equation}

\begin{align}
\ddot{a}  &  =\frac{V_{0}\left(  \sqrt{3}+\sqrt{3+2\omega_{BD}}\right)
}{\sqrt{3+2\omega_{BD}}}a^{-2+\frac{6}{3+\sqrt{3\left(  3+2\omega_{BD}\right)
}}}\Xi^{\frac{1+\sqrt{3\left(  3+2\omega_{BD}\right)  }}{2}}-\frac
{2\sqrt{3\left(  3+2\omega_{BD}\right)  }}{3+\sqrt{3\left(  3+2\omega
_{BD}\right)  }}\frac{\dot{a}^{2}}{a}\nonumber\\
&  -\frac{27\left(  \sqrt{\left(  3+2\omega_{BD}\right)  }-\sqrt{3}%
+\omega_{BD}\left(  \sqrt{3}\left(  33+\omega_{BD}\right)  +18\sqrt{\left(
3+2\omega_{BD}\right)  }\right)  \right)  }{\left(  3+2\omega_{BD}\right)
^{\frac{3}{2}}\left(  3+\sqrt{3\left(  3+2\omega_{BD}\right)  }\right)  }%
\dot{a}\frac{\dot{\Xi}}{\Xi}\nonumber\\
&  +\frac{\omega_{BD}\left(  \sqrt{3}\left(  3+\omega_{BD}\right)
+\sqrt{\left(  3+2\omega_{BD}\right)  }\right)  }{4\left(  3+2\omega
_{BD}\right)  ^{\frac{3}{2}}}a\left(  \frac{\dot{\Xi}}{\Xi}\right)  ^{2}.
\end{align}

From the second-order differential equation (\ref{s11}) we derive the
conservation law%
\begin{equation}
\Xi^{\frac{\sqrt{3\left(  3+2\omega_{BD}\right)  }-3+\omega_{BD}\left(
\sqrt{3\left(  3+2\omega_{BD}\right)  }-2\right)  }{2\left(  3+2\omega
_{BD}\right)  }}\dot{\Xi}=\tilde{I}_{1}%
\end{equation}
which is a Noetherian conservation law related with the vector field $\bar
{X}_{3}$, that is, $\tilde{I}_{1}\simeq I\left(  \bar{X}_{3}\right)  \,$.

Thus, for the scalar field $\Xi$ it follows that the closed-form solution is
given by%
\begin{equation}
\Xi\left(  t\right)  =\Xi_{0}\left(  3+\sqrt{3\left(  3+2\omega_{BD}\right)
}+\omega_{BD}\left(  2+\sqrt{\left(  3+2\omega_{BD}\right)  }\right)
t\right)  ^{\frac{2\left(  3+2\omega_{BD}\right)  }{3+\sqrt{3\left(
3+2\omega_{BD}\right)  }+\omega_{BD}\left(  2+\sqrt{3\left(  3+2\omega
_{BD}\right)  }\right)  }}.
\end{equation}
Finally, for the scalar factor we determine the exact solution
\begin{equation}
a\left(  t\right)  =t^{\frac{5+4\omega_{BD}+\sqrt{3\left(  3+2\omega
_{BD}\right)  }}{\left(  4+3\omega_{BD}\right)  }}~.
\end{equation}%
\begin{equation}
V_{0}=\frac{2\left(  3+2\omega_{BD}\right)  }{\left(  4+3\omega_{BD}\right)
^{2}}\left(  13+9\omega_{BD}+\sqrt{3\left(  3+2\omega_{BD}\right)  }\Xi
_{0}^{-\frac{1+\sqrt{3\left(  3+2\omega_{BD}\right)  }}{2}}\right)  \text{. }%
\end{equation}
Consequently, the Hubble function and the effective equation of state
parameter are derived
\begin{equation}
H\left(  t\right)  =\frac{5+4\omega_{BD}+\sqrt{3\left(  3+2\omega_{BD}\right)
}}{\left(  4+3\omega_{BD}\right)  }\frac{1}{t}~,
\end{equation}%
\begin{equation}
w_{eff}\left(  t\right)  =-1+\frac{22\left(  4+3\omega_{BD}\right)  }{3\left(
5+4\omega_{BD}+\sqrt{3\left(  3+2\omega_{BD}\right)  }\right)  }.
\end{equation}

Therefore, the exact solution describes an accelerated Universe when
\begin{equation}
-\frac{1}{16}\left(  \sqrt{33}+17\right)  <\omega_{BD}<\frac{6}{841}\left(
\sqrt{22}-188\right)  .
\end{equation}

\section{Conclusions}

\label{conc}

The Noether symmetry analysis is a powerful method for the study of nonlinear
dynamical systems with a variational principle. The symmetry analysis has been
widely applied in gravitational systems for the construction of conservation
laws and the study of various cosmological models.

In this study we applied the Noether symmetry analysis in order to study the
nonlinear field equations for a two-scalar field cosmological model in a
spatially flat FLRW geometry. The gravitational theory is defined in the
Jordan frame, where one of the scalar fields is the Brans-Dicke field and the
second scalar field is minimally coupled gravity but nonminimally to the
Brans-Dicke field. This specific model has been proposed before as the
analogue in the Jordan frame for the Chiral model which generates the hyperinflation.

The cosmological model possesses three arbitrary parameters, namely the
$\left(  \omega_{BD},\kappa,\lambda\right)  $. From the application of
Noether's theorem it was found that for specific sets of the variables
$\left(  \kappa\left(  \omega_{BD}\right)  ,\lambda\left(  \omega_{BD}\right)
\right)  $ the field equations admit additional conservation laws such that
the field equations constitute a super-integrable dynamical system. For the
these cases, with the use of the normal coordinates we were able to simplify
the field equations and to write the closed-form and exact solutions. The
analysis of the solutions gives constraint for the free parameter,
$\omega_{BD}$, such that the hyperbolic inflationary solution be recovered.

In a future work we plan to investigate further these super-integrable models
and in particular we plan \ to solve the Wheeler-DeWitt equation of quantum
cosmology and compare the semiclassical limit in the Jordan and in
the\ Einstein frames.

\end{document}